\documentclass[12pt]{amsart}
\usepackage{amsmath}
\usepackage{latexsym}
\usepackage{amsfonts}
\usepackage{amssymb}
\usepackage{color}
\usepackage[english]{babel}

\newtheorem{proposition}{Proposition}

\newtheorem{corollary}{Corollary}

\theoremstyle{definition}

\newtheorem{remark}{Remark}
\newtheorem{example}{Example}
\newtheorem{definition}{Definition}

\newcommand{\hi}{\mathcal{H}} 
\newcommand{\lh}{\mathcal{L(H)}} 
\newcommand{\sh}{\mathcal{S(H)}} 
\newcommand{\eh}{\mathcal{E(H)}} 

\newcommand{\ip}[2]{\left\langle\,#1\,|\,#2\,\right\rangle} 

\newcommand{\kb}[2]{|#1\,\rangle\langle\,#2|} 
\newcommand{\no}[1]{\left\|#1\right\|} 

\newcommand{\F}{\mathcal{F}}

\newcommand{\R}{\mathbb R}

\newcommand{\A}{\mathcal{A}}
\newcommand{\bor}[1]{\mathcal{B}({#1})}
\newcommand{\br}{\mathcal B(\mathbb R)}
\newcommand{\brrr}{\mathcal B(\R^3)}

\newcommand{\posr}{\mathcal{POS}_\R}
\newcommand{\posrrr}{\mathcal{POS}_{\R^3}}

\newcommand{\infeq}{\sim}

\begin{document}

\title[Position and momentum observables]{Position and momentum observables on $\R$ and on $\R^3$}

\author[Carmeli]{Claudio Carmeli}
\address{Claudio Carmeli, Dipartimento di Fisica, Universit\`a di Genova, and I.N.F.N., Sezione di Genova, Via
Dodecaneso 33, 16146 Genova, Italy}
\email{carmeli@ge.infn.it}

\author[Heinonen]{Teiko Heinonen}
\address{Teiko Heinonen, Department of Physics and Department of Mathematics, University of Turku,
FIN-20014 Turku, Finland}
\email{teiko.heinonen@utu.fi}

\author[Toigo]{Alessandro Toigo}
\address{Alessandro Toigo, Dipartimento di Fisica, Universit\`a di Genova, and I.N.F.N., Sezione di Genova, Via
Dodecaneso 33, 16146 Genova, Italy}
\email{toigo@ge.infn.it}


\maketitle{}

\begin{abstract}
We characterize all position and momentum observables on $\R$ and on $\R^3$. We
study some of their operational properties and discuss their covariant
joint observables.
\end{abstract}

\section{Introduction}

In the traditional presentation of quantum mechanics, 
observables are represented by selfadjoint operators
 or, equivalently, by spectral measures. It is widely recognized that
 this concept is too narrow.
Indeed, spectral measures correspond to measurements with perfect
accuracy, never found in real experiments. A less restrictive mathematical
formulation of a quantum mechanical observable is a normalized
positive operator measure. This generalization allows one, among
many other things, to describe measurements with limited accuracy.
(For a review of positive operator measures in quantum mechanics,
see \cite{OQP}, \cite{Davies76}, \cite{Holevo82}.)

In this paper we take covariance and invariance with respect to
suitable symmetry groups as the defining properties of an
observable. For example, a position observable on $\R$ is defined to be
an observable which is covariant under space translations and invariant
under momentum boosts. We characterize all position and momentum
observables on $\R$ and on $\R^3$.

In Section \ref{R} we study position and momentum observables on
$\R$. In Subsection \ref{Def1} the definitions are given   and in Subsection \ref{Structure} we characterize the
structure of
position and momentum observables. In Subsection
\ref{State} we investigate the ability of a position observable to
discriminate states, that is, the state distinction power.
Another relevant property, the limit of resolution, is studied in
Subsection \ref{Resolution}. In Subsection \ref{Joint} we consider a
covariant joint observable of position and momentum  in phase space
and derive a lower bound for the product of their limits of resolution. Section \ref{R3} is devoted to
studying position and momentum observables in $\R^3$. The corresponding
definitions are formulated in Subsection \ref{Def3}, and a complete
classification of position and momentum observables in $\R^3$ is given
in Subsection \ref{Structure3}.

Concluding this section we fix some notations. Let $\hi$ be a
complex separable Hilbert space and $\lh$ the set of bounded
operators on $\hi$. A positive operator $T\in\lh$ of trace one is
called a \emph{state} and the set of all states is denoted by $\sh$. A
positive operator $A$ bounded from above by the unit operator $I$ is called
an \emph{effect} and the set of all effects is denoted by $\eh$. Way say
that the null operator $O$ and the unit operator $I$ are \emph{trivial
effects}. Let $\Omega$ be a nonempty set and $\A$ a $\sigma$-algebra of
subsets of $\Omega$. A set function $E:\A\to\lh$ is an
\emph{operator measure}, if it is $\sigma$-additive with respect
to the strong (or \cite[p. 318]{DunSch}, equivalently, weak) operator topology. 

We call an operator valued measure $E$ an \emph{observable} if
$E(X)\in\eh$ for all $X\in\A$ and $E(\Omega)=I$. If an
observable $E$ has projections as its values,
that is, $E(X)^*=E(X)=E(X)^2$ for all $X\in\A$, it is called a
\emph{sharp observable}. For an observable
$E:\A\to\lh$ and a state $T\in\sh$, we let $p^E_T$ denote the
probability measure on $\Omega$ defined by
$$
p^E_T(X)=tr[TE(X)],\ X\in\A.
$$
The number $p^E_T(X)$ is interpreted as the probability of having an
outcome in $X$ when the system is in the state $T$ and the observable
$E$ is measured.

We denote by $\bor{\R^n}$ the Borel $\sigma$-algebra of $\R^n$.
The Fourier transform of any $f\in
L^1(\R^n)$ is denoted by $\hat{f}$. 
We set also $\hat{f}=\F(f)$ to denote the Fourier-Plancherel transform of any $f\in L^2(\R^n)$ and 
$\hat{\mu}=\F(\mu)$ is the Fourier-Stieltjes transform of any complex Borel
measure $\mu$ on $\R^n$. 

\section{Position and momentum observables on $\R$}\label{R}

\subsection{Definitions}\label{Def1}

Let us consider a non-relativistic particle living in the
one-dimensional space $\R$ and fix $\hi=L^2(\R)$. Let $U$ and $V$ be
the one-parameter unitary representations on $\hi$ related to the
groups of space translations and momentum boosts. They act on
$\varphi\in\hi$ as
\begin{eqnarray*}
\left[U(q)\varphi\right](x) &=& \varphi(x-q), \\
\left[V(p)\varphi\right](x) &=& e^{ipx}\varphi(x).
\end{eqnarray*}
Let $P$ and $Q$ be the selfadjoint
operators generating $U$ and $V$, that is, $U(q)=e^{-iqP}$ and
$V(p)=e^{ipQ}$ for every $q,p\in\R$. We denote by $\Pi_P$ and $\Pi_Q$
the spectral decompositions of the operators $P$ and $Q$, respectively. They have the form 
\begin{eqnarray*}
&& \left[\Pi_Q(X)\varphi\right](x) = \chi_X(x)\varphi(x),\\
&& \Pi_P(X) = \F^{-1}\Pi_Q(X)\F.
\end{eqnarray*}
The sharp observable $\Pi_Q$ has the property that, for all $q,p\in\R$ and $X\in\br$,
\begin{eqnarray}
U(q)\Pi_Q(X)U(q)^* &=& \Pi_Q(X+q),\label{covQ} \\
V(p)\Pi_Q(X)V(p)^* &=& \Pi_Q(X)\label{invQ}.
\end{eqnarray}
Equation (\ref{covQ}) means that $\Pi_Q$ is covariant under
translations whereas (\ref{invQ})
 shows that $\Pi_Q$ is invariant under momentum boosts. Hence, these
 relations suggest to call $\Pi_Q$ a position observable. As the
 kinematical meaning of the observable $\Pi_Q$ is solely
in the relations (\ref{covQ}) and (\ref{invQ}), we take these
symmetry properties as the definition of a general position
observable. The observable $\Pi_Q$ is called the \emph{canonical position observable}.

\begin{definition}
An observable $E:\br\to\lh$ is a \emph{position
observable on $\R$} if, for all $q,p\in\R$ and $X\in\br$,
\begin{eqnarray}
U(q)E(X)U(q)^* &=& E(X+q),\label{cov}\\
V(p)E(X)V(p)^* &=& E(X).\label{inv}
\end{eqnarray}
We will denote by $\posr$ the set of all position observables on
$\R$.
\end{definition}

In an analogous way we define a momentum observable to be an
observable which is covariant under momentum boosts and invariant
under translations.

\begin{definition}
An observable $F:\br\to\lh$ is a \emph{momentum
observable on} $\R$ if, for all $q,p\in\R$ and $X\in\br$,
\begin{eqnarray}
U(q)F(X)U(q)^* &=& F(X),\label{invF}\\
V(p)F(X)V(p)^* &=& F(X+p).\label{covF}
\end{eqnarray}
\end{definition}

Since $\F U(q) = V(-q) \F$ and $\F V(p) = U(p) \F$, the sharp observable $\Pi_P =
\F^{-1}\Pi_Q \F$ satisfies (\ref{invF}) and (\ref{covF}). It is called the \emph{canonical momentum observable}. 
Moreover, an observable
$E$ is a position observable if and only if $\F^{-1}E \F$ is a momentum observable.
Therefore, in the following we will restrict ourselves to the study of position
observables, the results of Sections \ref{Structure}, \ref{State} and
\ref{Resolution} being easily converted to the case of momentum
observables.

\begin{remark}
In some articles an observable $E:\br\to\lh$ satisfying the covariance condition
(\ref{covQ}) (and not necessarily the invariance condition
(\ref{invQ})) is called a (generalized) position observable. In this
paper we say that such an observable is a \emph{localization
  observable}. These are characterized in \cite{CDT2},~\cite{Artevo}. In
Subsection \ref{Structure} it is shown, especially, that every
position observable is commutative. However, there exist
noncommutative localization observables. Hence, the set $\posr$ is a
proper subset of all localization observables.
\end{remark}

\subsection{The structure of position observables}\label{Structure}

Let $\rho:\br\to [0,1]$ be a probability measure. For any
$X\in\br$, the map $q\mapsto\rho(X-q)$ is bounded and measurable,
and hence the equation
\begin{equation}\label{pos1}
E_{\rho}(X)=\int \rho(X-q)\ d\Pi_Q(q)
\end{equation}
defines a bounded positive operator.
The map
$$
\br\ni X\mapsto E_{\rho}(X)\in\lh
$$
is an observable. It
is straightforward to verify that the observable $E_{\rho}$
satisfies the covariance condition (\ref{cov}) and the invariance condition
(\ref{inv}), hence it is a position observable on $\R$. Denote by
$\delta_t$ the Dirac measure concentrated at $t$. The observable
$E_{\delta_0}$ is
the canonical position observable $\Pi_Q$. We may also write
\begin{equation}\label{pos2}
\Pi_Q(X)=\int \delta_0(X-q)\ d\Pi_Q(q)
\end{equation}
and comparing (\ref{pos1}) to (\ref{pos2}) we note that $E_{\rho}$ is
obtained when the sharply concentrated Dirac measure $\delta_0$ is replaced by
the probability measure $\rho$. The observable $E_{\rho}$ admits an
interpretation as an imprecise, or fuzzy, version of the canonical position observable $\Pi_Q$. (See \cite{AliEmch}, 
\cite{AliDoe}, \cite{AliPru} for further details.)

\begin{proposition}\label{struc1}
Any position observable $E$ on $\R$ is of the form $E=E_{\rho}$ for some probability measure
$\rho:\br\to [0,1]$.
\end{proposition}
The proof of Proposition \ref{struc1} is given in Appendix \ref{proof1}.

Besides covariance (\ref{covQ}) and invariance (\ref{invQ}),
the canonical position observable $\Pi_Q$ has still more symmetry
properties. Namely, let $\R_+$ be the set of positive real numbers
regarded as a multiplicative group. It has a family of unitary
representations $\left\{ A_t \mid t\in \R \right\}$ acting on
$\hi$, and given by
$$
[A_t(a)f](x)=\frac{1}{\sqrt{a}}f\left(a^{-1}(x-t)+t\right).
$$
It is a direct calculation to verify that for all
$a\in\R_+,X\in\br$,
$$
A_0(a)\Pi_Q(X) A_0(a)^*=\Pi_Q(aX).
$$
We adopt the following terminology.
\begin{definition}
We say that an observable $E:\br\to\lh$ is \emph{covariant under
dilations} if there exists a unitary representation $A$ of $\R_+$
such that for all $a\in\R_+$ and $X\in\br$,
\begin{equation} \label{e_dilat}
A(a)E(X)A(a)^*=E(aX).
\end{equation}
\end{definition}

The canonical position observable $\Pi_Q$ is not the only position
observable which is covariant under dilations. An observable
$E_{\delta_t}, t\in\R,$ is a translated version of
$\Pi_Q$, namely, for any $X\in\br$,
$$
E_{\delta_t}(X)=\Pi_Q(X-t)=U(t)^*\Pi_Q(X)U(t).
$$
Since $A_{-t}(a) = U(t)^* A_{0}(a) U(t)$, the observable $E_{\delta_t}$
is covariant under dilations, with, for example,
$A = A_{-t}$.

\begin{proposition}\label{dilat}
Let $E$ be a position observable on $\R$. The following conditions are
equivalent:
\begin{itemize}
\item[(a)] $E$ is covariant under dilations;
\item[(b)] $\no{E(U)}=1$ for every nonempty open set $U\subset\R$;
\item[(c)] $E=E_{\delta_t}$ for some $t\in\R$;
\item[(d)] $E$ is a sharp observable.
\end{itemize}
\end{proposition}

\begin{proof}
Let $E$ be covariant under dilations. In a similar way as in
\cite[Lemma 3]{Cast} it can be shown that $\no{E(U)}=1$ for all
nonempty open sets $U$, and so, (a) implies (b). Assume then that (b) holds.
For any nonempty open set $U$ we get
\begin{equation}\label{norm1}
1=\no{E(U)}={\rm ess\,sup}_{x\in\R} \rho(x+U).
\end{equation}
It follows that ${\rm supp}(\rho)$ contains only one point.
Indeed, assume on the contrary that ${\rm supp}(\rho)$
contains two points $x_1\neq x_2$ and denote $U=\{x\in\R |
|x|<\frac{1}{4}|x_1-x_2|\}$. Since $x_1+U$ and $x_2+U$ are
neighborhoods of $x_1$ and $x_2$, respectively, we have
$m_i:=\rho(x_i+U)>0$ for $i=1,2$. Then, for any $x\in\R$,
$\rho(x+U)\leq 1-{\rm min}(m_1,m_2)$. This is in contradiction
with (\ref{norm1}).  Hence, (b) implies (c). As previously
mentioned, (c) implies (a). Clearly, (c) also implies (d). Since
(d) implies (b) the proof is complete.
\end{proof}

The dilation covariance means that the observable in question has
no scale dependence. A realistic position measurement apparatus has a limited
accuracy and hence it cannot define a position observable which is
covariant under dilations. Thus, sharp position
observables are not suitable to describe nonideal situations.

\begin{remark}\label{r_dilation}
If $A$ is a unitary representation of $\mathbb{R}_+$ satisfying
Eq.~(\ref{e_dilat}) with $E=E_{\delta_t}$, then there exists a
measurable function $\beta : \R \longrightarrow \mathbb{T}$
($\mathbb{T}=$ the complex numbers of modulus $1$) such
that
\begin{equation*}
\left[A(a)f\right](x) = \frac{1}{\sqrt{a}} \beta (x+t)
\overline{\beta \left(a^{-1}(x+t)\right)}
f\left(a^{-1}(x+t)-t\right).
\end{equation*}
In particular, $A$ is equivalent to $A_{-t}$. See Appendix
\ref{supplement} for more details.
\end{remark}

\subsection{State distinction power of a position observable}\label{State}

\begin{definition}
Let $E_1$ and $E_2$ be observables on $\R$. The \emph{state
  distinction power} of $E_2$ is greater than or equal to $E_1$ if for
all $T,T'\in\sh$,
$$
p^{E_2}_T=p^{E_2}_{T'}\ \Rightarrow p^{E_1}_T=p^{E_1}_{T'}.
$$
In this case we denote $E_1 \sqsubseteq E_2$. If $E_1 \sqsubseteq
E_2 \sqsubseteq E_1$ we say that $E_1$ and $E_2$ are
\emph{informationally equivalent} and denote $E_1\infeq E_2$.
 If $E_1 \sqsubseteq E_2$ and $E_2 \not\sqsubseteq E_1$, we write $E_1 \sqsubset E_2$.
\end{definition}

\begin{example}\label{trivial1}
An observable $E:\br\to\lh$ is \emph{trivial} if
$p^E_T=p^E_{T'}$ for all states $T,T'\in\sh$. This implies that a
trivial observable $E$ is of the form $E(X)=\lambda(X) I,X\in\br$,
for some probability measure $\lambda$. The state distinction
power of any observable $E'$ is greater than or equal to that
of the trivial observable $E$. Clearly there is  no trivial position observable on $\R$.
\end{example}

\begin{example}
An observable $E:\br\to\lh$ is called \emph{informationally
complete} if $p^E_T\neq p^E_{T'}$ whenever $T\neq T'$. The state
distinction power of an informationally complete observable is
greater than or equal to that of any other observable $E_1$ on $\R$. It is
easy to see that there is no informationally complete position
observable. Namely, let $\psi$ be a unit vector, $p\neq 0$ a real
number, and denote $\psi'= V(p)\psi$. Then the states
$T=\kb{\psi}{\psi}$ and $T'=\kb{\psi'}{\psi'}$ are different
but for any position observable $E_{\rho}$,
$p^{E_{\rho}}_T=p^{E_{\rho}}_{T'}$ since $V(p)$ commutes with all the effects $E_{\rho}(X), X\in\br$.
\end{example}

We will next think of $\infeq$ as a relation on the set $\posr$. The
relation $\infeq$ is clearly reflexive, symmetric and transitive,
and hence it is an equivalence relation. We denote the
equivalence class of a position observable $E$ by $[E]$ and
the space of equivalence classes as $\posr/\infeq$. The relation
$\sqsubseteq$ induces a partial order in the set $\posr/\infeq$ in
a natural way.

Let $E_{\rho}$ be a position observable and $T$ a state. The
probability measure $p^{E_{\rho}}_T$ is the convolution of
the probability measures $p^{\Pi_Q}_T$ and $\rho$,
\begin{equation}\label{convol}
p^{E_{\rho}}_T=p^{\Pi_Q}_T\ast\rho.
\end{equation}
It is clear from (\ref{convol}) that for all $T,T'\in\sh$,
$$
p^{\Pi_Q}_T=p^{\Pi_Q}_{T'}\ \Rightarrow
p^{E_{\rho}}_T=p^{E_{\rho}}_{T'},
$$
and hence $E_{\rho}\sqsubseteq \Pi_Q$. We conclude that
$[\Pi_Q]$ is the only maximal element of the partially
ordered set $\posr/\infeq$.

It is shown in \cite[Prop. 5]{fuzzy} that a position observable
$E_\rho$ belongs to the maximal equivalence class $[\Pi_Q]$
 if and only if
${\rm supp}\left(\widehat{\rho}\right)=\R$. The following proposition
characterizes the equivalence classes completely.

\begin{proposition}
Let $\rho_1,\rho_2$ be probability measures on $\R$ and $E_{\rho_1}, E_{\rho_2}$
the corresponding position observables. Then
\begin{equation}
E_{\rho_1}\sqsubseteq E_{\rho_2} \Leftrightarrow
{\rm supp}\left(\widehat{\rho_1}\right)\subseteq {\rm supp}
\left(\widehat{\rho_2}\right).
\end{equation}
\end{proposition}

\begin{proof}
Taking the Fourier transform of Eq.~(\ref{convol}), we get
\begin{equation} \label{e_Fourier}
\F(p_T^{E_\rho}) = \F(p_T^{\Pi_Q}) \F(\rho).
\end{equation}
Since the Fourier transform is injective, it
is clear from the above relation that ${\rm supp} \left( \widehat{\rho}_1 \right)  \subseteq
{\rm supp} \left( \widehat{\rho}_2 \right) $ implies
$E_{\rho_1}\sqsubseteq E_{\rho_2}$.

Conversely, suppose ${\rm supp} \left( \widehat{\rho}_1 \right)
\nsubseteq {\rm supp }\left( \widehat{\rho}_2 \right) $. As
$\widehat{\rho}_i$, $i=1,2$, are continuous functions and $\widehat{\rho}_i
\left(\xi \right) = \overline{\widehat{\rho}_i \left(-\xi \right)}$,
there exists a closed interval $[2a,2b]$, with $0\leq a < b$, such
that $[2a,2b]\cup [-2b,-2a]\subseteq {\rm supp \left( \widehat{\rho}_1
\right)}$ and $\left( [2a,2b]\cup [-2b,-2a] \right) \cap {\rm
supp} \left( \widehat{\rho}_2 \right) = \emptyset$. Define the
functions
\begin{eqnarray*}
h_1 &=& \frac{1}{\sqrt{2(b-a)}} \left(\chi_{[a,b]} -
\chi_{[-b,-a]}
\right), \\
h_2 &=& \frac{1}{\sqrt{2(b-a)}} \left(\chi_{[a,b]} +
\chi_{[-b,-a]} \right),
\end{eqnarray*}
and for $i=1,2$, denote
\begin{equation*}
h_i^*(\xi) := \overline{h_i(-\xi)}.
\end{equation*}
Define
\begin{equation*}
f_i = \mathcal{F}^{-1}\left( h_i \right),
\end{equation*}
and let $T_i$ be the one-dimensional projection $\left|f_i\right\rangle\left\langle f_i \right|$.
We then have
\begin{equation*}
dp_{T_i}^{\Pi_Q} (x) = \left\vert f_i(x) \right\vert ^2 dx =
\left\vert \left(\mathcal{F}^{-1} h_i \right)(x) \right\vert ^2 dx =
\mathcal{F}^{-1}\left( h_i*h_i^* \right)(x) dx,
\end{equation*}
and
\begin{equation*}
\begin{split}
\F(p_{T_i}^{\Pi_Q}) & = \mathcal{F}\mathcal{F}^{-1}\left(
h_i*h_i^* \right)=  h_i*h_i^* \\
&= \frac{1}{2(b-a)} \Bigl(2\chi_{[a,b]}*\chi_{[-b,-a]} +
(-1)^i\chi_{[-b,-a]}*\chi_{[-b,-a]}  \\
&\quad + (-1)^i\chi_{[a,b]}*\chi_{[a,b]}\Bigr).
\end{split}
\end{equation*}
Since
\begin{eqnarray*}
 {\rm supp}\left( \chi_{[a,b]}*\chi_{[-b,-a]}\right) & = & [a-b,b-a],\\
 {\rm supp}\left( \chi_{[a,b]}*\chi_{[a,b]}\right) & = & [2a,2b], \\
{\rm supp} \left(\chi_{[-b,-a]}*\chi_{[-b,-a]}\right) & = & [-2b,-2a],
\end{eqnarray*}
an application of (\ref{e_Fourier}) shows that
\begin{eqnarray*}
\F(p_{T_1}^{E_{\rho_1}}) &\neq& \F(p_{T_2}^{E_{\rho_1}}), \\
\F(p_{T_1}^{E_{\rho_2}}) &=& \F(p_{T_2}^{E_{\rho_2}}),
\end{eqnarray*}
or in other words, $E_{\rho_1} \not \sqsubseteq E_{\rho_2}$.
\end{proof}

\begin{remark}
It follows from the above proposition that $E_1 \sqsubset E_2
\Leftrightarrow {\rm supp}\left(\widehat{\rho_1}\right)\subset {\rm supp}
\left(\widehat{\rho_2}\right)$, and hence the set $\posr/\infeq$ has
no minimal element. Indeed, if $\rho_2$ is a probability measure, there always
exists a probability measure $\rho_1$ such that ${\rm supp}\left(\widehat{\rho_1}\right)
\subset {\rm supp}\left(\widehat{\rho_2}\right)$. In fact, since
$\widehat{\rho_2}$ is continuous, $\widehat{\rho_2}(\xi)=
\overline{\widehat{\rho_2}(-\xi)}$ and $\widehat{\rho_2}(0)=
\rho_2 (\mathbb{R})\neq 0$, there exists $a>0$ such that the closed
interval $[-a,a]$ is strictly contained in
${\rm supp}\left(\widehat{\rho_2}\right)$. If we define
$h=\frac{1}{\sqrt{a}} \chi_{\left[-\frac{a}{2},\frac{a}{2}\right]}$,
$f=\mathcal{F}^{-1}h$, then $d\rho_1 (x):= \vert f(x) \vert^2 dx$
is a probability measure, and ${\rm supp}\left(\widehat{\rho_1}\right)=
{\rm supp}\left(h*h\right)=[-a,a]$.
\end{remark}

\subsection{Limit of resolution of a position observable}\label{Resolution}

Let $\Pi:\br\to\lh$ be a sharp observable. For any nontrivial projection
$\Pi(X)$, there exist states
$T,T'\in\sh$ such that $p^\Pi_T(X)=1$ and $p^\Pi_{T'}(X)=0$. We may say
that $\Pi(X)$ is a \emph{sharp property} and it is \emph{real} in the
state $T$.

In general, an observable $E$ has effects as its values which are not
projections and, hence, not sharp properties.
An effect $B\in\eh$ is called \emph{regular} if
its spectrum extends both above and below $\frac{1}{2}$. This
means that there exist states $T,T'\in\sh$ such that
${\rm tr}[TB]>\frac{1}{2}$ and ${\rm tr}[T'B]<\frac{1}{2}$. In this
sense regular effects can be seen as \emph{approximately
realizable properties}, see \cite[II.2.1]{OQP}. The observable $E$
is called regular if all the nontrivial effects $E(X)$ are
regular.

It is shown in \cite[Prop. 4]{fuzzy} that if a probability measure
$\rho$ is absolutely continuous with respect to the Lebesgue measure,
then the position observable $E_{\rho}$ is not regular.  Here we modify the
notion of regularity to get a quantification of sharpness, or
resolution, of position observables.

For any $x\in\R, r\in\R_+$, we denote the interval
$[x-\frac{r}{2},x+\frac{r}{2}]$ by $I_{x;r}$. We also denote
$I_r=I_{0;r}$. Let $E:\br\to\lh$ be an observable and
$\alpha>0$. We say that $E$ is $\alpha$-\emph{regular} if all the
nontrivial effects $E(I_{x;r})$, $x\in\R$, $r\geq\alpha$, are
regular.

\begin{definition}
Let $E:\br\to\lh$ be an observable. We denote
$$
\gamma_E=\inf \{\alpha>0 \mid \textrm{$E$ is $\alpha$-regular}\}
$$
and say that $\gamma_E$ is the \emph{limit of resolution of $E$}.
\end{definition}

It follows directly from definitions that the limit of resolution
of a regular observable is 0. Especially, the limit of resolution of
canonical position observables is 0.

\begin{example}\label{trivial2}
Let $E$ be a trivial observable (see Example \ref{trivial1}). For any
$X\in\br$, we have either $E(X)\geq\frac{1}{2}I$ or
$E(X)\leq\frac{1}{2}I$. Hence, $\gamma_E=\infty$.
\end{example}

\begin{proposition} \label{reg2}
A position observable $E_{\rho}$ is $\alpha$-regular if and only
if
\begin{equation}\label{areg}
{\rm ess\,sup}_{x\in\R}\rho(I_{x,\alpha})>\frac{1}{2}.
\end{equation}
\end{proposition}

\begin{proof}
An effect $E_{\rho}(X)$ is regular if and only if
$\no{E_{\rho}(X)}>\frac{1}{2}$ and
$\no{E_{\rho}(\R\setminus X)}>\frac{1}{2}$. Since the norm of the
multiplicative operator $E_{\rho}(X)$ is ${\rm ess\,sup}_{x\in\R}\rho(X-x)$,
we conclude that $E_{\rho}(X)$ is regular if and only if
$$
{\rm ess\,sup}_{x\in\R}\rho(X-x)>\frac{1}{2}\quad \textrm{ and}\quad {\rm ess\,inf}_{x\in\R}\rho(X-x)<\frac{1}{2}.
$$
Thus, $E_{\rho}$ is $\alpha$-regular
if and only if, for all $r\geq\alpha$,
$$
{\rm ess\,sup}_{x\in\R}\rho(I_{x;r})>\frac{1}{2}\quad \textrm{ and}\quad
{\rm ess\,inf}_{x\in\R}\rho(I_{x;r})<\frac{1}{2}.
$$
The second condition is always satisfied and the first is
equivalent to (\ref{areg}).
\end{proof}

\begin{corollary}
A position observable
$E_{\rho}$ has a finite limit of resolution and
\begin{equation}
\gamma_{E_{\rho}}=\inf\{\alpha > 0 \mid
{\rm ess\,sup}_{x\in\R}\rho(I_{x;\alpha})>\frac{1}{2}\}.
\end{equation}
\end{corollary}

\begin{example}\label{Gaussian}
Let us consider the case in which the probability
measure has Gaussian distribution, that is,
\begin{equation*}
d \rho(x) =
\frac{1}{\sigma\sqrt{2\pi}}e^{-\frac{(x-\bar{x})^2}{2\sigma^2}} dx.
\end{equation*}
By Proposition \ref{reg2} the position observable $E_{\rho}$ is
$\alpha$-regular if, for each $r\geq\alpha$,
\begin{equation*}
\frac{1}{2}< \int_{I_{\bar{x};r}}
\frac{1}{\sigma\sqrt{2\pi}}e^{-\frac{(x-\bar{x})^2}{2\sigma^2}} dx
=\frac{1}{\sqrt{2\pi}} \int_{-\frac{r}{2\sigma}}^{\frac{r}{2\sigma}}e^{-\frac{x^2}{2}} dx.
\end{equation*}

It follows that the limit of resolution $\gamma_{E_{\rho}}$ is
proportional to the standard deviation $\sigma$ and
$\gamma_{E_{\rho}}\approx 1.36\sigma$.
\end{example}

\subsection{Covariant joint observables of position and momentum observables}\label{Joint}

Let $E_1,E_2:\br\to\lh$ be two observables. An
observable $G:\bor{\R^2}\to\lh$ is their \emph{joint observable} if for all
$X,Y\in\br$,
\begin{eqnarray*}
E_1(X) &=& G(X\times\R),\\
E_2(Y) &=& G(\R\times Y).
\end{eqnarray*}
In this case $E_1$ and $E_2$ are the \emph{margins} of $G$.

For all $(q,p)\in\R^2$, we denote
$$
W(q,p)=e^{-iqP+ipQ}=e^{iqp/2}U(q)V(p).
$$
The mapping $W:(q,p)\mapsto W(q,p)$ is an irreducible projective
representation of the phase space translation group $\R^2$ in $\hi$. An
observable $G:\bor{\R^2}\to\lh$ is called a \emph{covariant phase space observable} if for all
$(q,p)\in\R^2$ and $Z\in\bor{\R^2}$,
$$
W(q,p)G(Z)W(q,p)^*=G(Z+(q,p)).
$$
It is proved in \cite[III.A.]{CDT1} that all covariant phase space
observables are of the form
\begin{equation}\label{phase}
G_T(Z)=\frac{1}{2\pi}\int_Z W(q,p)TW(q,p)^*\ dqdp
\end{equation}
for some $T\in\sh$.

The margins of a covariant phase space observable $G_T$ are position
and momentum observables. Indeed, let $\sum_i \lambda_i
\kb{\varphi_i}{\varphi_i}$ be the spectral decomposition of $T$. A straightforward calculation shows that
\begin{equation}
G_T(X\times \R)=\int \rho(X-q)\ d\Pi_Q(q)=E_{\rho}(X),
\end{equation}
 where $d\rho(q)=e(q)dq$ and $e(q)=\sum_i \lambda_i
|\varphi_i(-q)|^2$. Similarly,
\begin{equation}
G_T(\R\times Y)=\int \nu(Y-p)\ d\Pi_P(p) = F_{\nu}(Y),
\end{equation}
 where $d\nu(p)=f(p)dp$ and $f(p)=\sum_i \lambda_i
|\widehat{\varphi_i}(-p)|^2$.

The following proposition is a direct consequence of the previously
mentioned results.

\begin{proposition}
A position observable $E_{\rho}$ [a momentum observable $F_{\nu}$] is
a margin of a phase space observable if and only if the probability
measure $\rho$ [prob. measure $\nu$] is
absolutely continuous with
respect to the Lebesgue measure.
\end{proposition}

As noted in Example \ref{Gaussian}, the limit of resolution of a position
observable $E_{\rho}$ with $\rho$ having Gaussian distribution is
proportional to the standard deviation $\sigma$ of $\rho$. This shows,
in particular, that there exists a position observable which is
a margin of a phase space observable and which has an arbitrary
small positive limit of resolution. However, we next show that if
position and momentum observables have a covariant phase space
observable as their joint observable, then the product of limit of resolutions has a lower bound.

\begin{proposition}\label{up}
Let $E_{\rho}$ be a position observable and $F_{\nu}$ a momentum observable. If
$E_{\rho}$ and $F_{\nu}$ have a covariant phase space observable as their joint
observable, then
\begin{equation}\label{ineq_up}
\gamma_{E_{\rho}}\cdot\gamma_{F_{\nu}}\geq 3-2\sqrt 2.
\end{equation}
\end{proposition}

\begin{proof}
Since $E_{\rho}$ and $F_{\nu}$ have a covariant phase space observable as a
joint observable there is a vector valued function $\theta\in
L^2(\R,\hi)$ such that $d\rho(q)=\|\theta(q)\|_{\hi}^2 dq$ and
$d\nu(p)=\|\hat{\theta}(p)\|_{\hi}^2 dp$.

By Proposition \ref{reg2} the observable $E_\rho$ is $\alpha$-regular if and only if
$$
{\rm ess\,sup}_{x\in\mathbb{R}} \rho \left( I_{x;\alpha} \right)
> 1/2.
$$
Since the map $x\longmapsto \rho \left( I_{x;\alpha}
\right)$ is continuous, this is equivalent to
\begin{equation*}
{\rm sup}_{x\in \mathbb{R}} \rho \left( I_{x;\alpha} \right) =
{\rm sup}_{x\in \mathbb{R}} \int_{I_{x;\alpha}} \left\| \theta (x)
\right\|_{\hi}^2 dx > 1/2.
\end{equation*}
By the same argument, $F_\nu$ is $\beta$-regular if and only if
\begin{equation*}
{\rm sup}_{\xi\in \mathbb{R}} \nu \left( I_{\xi;\beta} \right) =
{\rm sup}_{\xi\in \mathbb{R}} \int_{I_{\xi;\beta}} \|
\hat{\theta} (\xi) \|_{\hi}^2 d\xi > 1/2.
\end{equation*}
Using \cite[Theorem 2]{DonSta} extended to the case of vector valued
functions, we find
\begin{equation*}
\alpha\cdot \beta \geq 3-2\sqrt{2},
\end{equation*}
and hence (\ref{ineq_up}) follows.
\end{proof}

\section{Position and momentum observables on $\R^3$}\label{R3}

\subsection{Definitions}\label{Def3}

In this section $\hi=L^2(\R^3)$.
Let $Q_i,i=1,2,3$, denote the multiplication operator on $\hi$ given
by $\left[Q_if\right](\vec{x})=x_if(\vec{x})$, where $x_i$ is the $i$th component
of $\vec{x}$. By $P_i$ we mean the operator $\F^{-1}Q_i\F$ and we
denote $\vec{Q}=(Q_1,Q_2,Q_3)$, $\vec{P}=(P_1,P_2,P_3)$. The space
translation group $\R^3$ has a unitary representation
$U(\vec{q})=e^{-i\vec{q}\cdot\vec{P}}$ and similarly, the momentum boost
group has a representation $V(\vec{p})=e^{i\vec{p}\cdot\vec{Q}}$. It
is an immediate observation that the sharp observables $\Pi_{\vec{Q}}$ and
$\Pi_{\vec{P}}$ on $\R^3$, associated to the representations $V$ and $U$,
respectively,
satisfy the obvious covariance and invariance conditions, analogous to
(\ref{cov})-(\ref{covF}).  Let $D$ be the
representation of the rotation group $SO(3)$ in $\hi$ defined as
$$
\left[D(R)f\right](\vec{x}) = f(R^{-1}\vec{x}).
$$
It is
straightforward to verify that the sharp observables $\Pi_{\vec{Q}}$ and
$\Pi_{\vec{P}}$ are covariant under rotations, that is, for all $R\in SO(3)$ and $X\in\bor{\R^3}$,
\begin{eqnarray*}
D(R)\Pi_{\vec{Q}}(X)D(R)^* &=& \Pi_{\vec{Q}}(RX), \\
D(R)\Pi_{\vec{P}}(X)D(R)^* &=& \Pi_{\vec{P}}(RX).
\end{eqnarray*}

These observations motivate to the following definitions.

\begin{definition}
An observable $E:\bor{\R^3}\to\lh$ is  a \emph{position observable on $\R^3$} if, for all
$\vec{q},\vec{p}\in\R^3$, $R\in SO(3)$ and $X\in\brrr$,
\begin{eqnarray}
U(\vec{q})E(X)U(\vec{q})^* &=& E(X+\vec{q}),\label{cov3}\\
V(\vec{p})E(X)V(\vec{p})^* &=& E(X),\label{inv3}\\
D(R)E(X)D(R)^* &=& E(RX). \label{rot3}
\end{eqnarray}
We will denote by $\posrrr$ the set of all position observables on
$\R^3$.
\end{definition}

\begin{definition}
An observable $F:\bor{\R^3}\to\lh$ is a \emph{momentum observable on $\R^3$} if, for all
$\vec{q},\vec{p}\in\R^3$, $R\in SO(3)$ and $X\in\brrr$,
\begin{eqnarray*}
U(\vec{q})F(X)U(\vec{q})^* &=& F(X),\\
V(\vec{p})F(X)V(\vec{p})^* &=& F(X+\vec{p}),\\
D(R)F(X)D(R)^* &=& F(RX).
\end{eqnarray*}
\end{definition}

\subsection{Structure of position observables on $\R^3$}\label{Structure3}

We say that a probability measure $\rho$ on $\R^3$ is rotation
invariant if for all $X\in\brrr$ and $R\in SO(3)$,
$$
(R\cdot\rho)(X):=\rho(R^{-1}X)\equiv\rho(X).
$$
The set of rotation invariant probability measures on $\R^3$ is
denoted by $M(\R^3)^+_{1,inv}$. Using the isomorphism $\R^3
\setminus \{0\} \simeq \R_+ \times S^2$ and the disintegration
of measures, the restriction of any measure
$\rho \in M(\R^3)^+_{1,inv}$ to the subset $\R^3
\setminus \{0\}$ can be written in the form
\begin{equation*}
\left.d\rho\right|_{\R^3
\setminus \{0\}} \left(\vec{r}\right) = d\rho_{\rm rad} \left( r\right)
d\rho_{\rm ang} \left(\Omega\right),
\end{equation*}
where $\rho_{\rm rad}$ is a finite measure on $\R_+$ with
$\rho_{\rm rad} (\R_+) = 1-\rho(\{0\})$, and $\rho_{\rm ang}$ is the
$SO(3)$-invariant measure on the sphere $S^2$ normalized to $1$.

Given a rotation invariant probability measure $\rho$, the formula
\begin{equation} \label{pos3}
E_{\rho}(X)=\int \rho(X-\vec{q})\ d\Pi_{\vec{Q}}(\vec{q}),\quad
X\in\brrr,
\end{equation}
defines a position observable on $\R^3$.

\begin{proposition}\label{struc3}
Any position observable $E$ on $\R^3$ is of the form $E=E_{\rho}$
for some $\rho\in M(\R^3)^+_{1,inv}$.
\end{proposition}

\begin{proof}
It is shown in Appendix \ref{proof1} that if $E$ satisfies
Eqs.~(\ref{cov3}), (\ref{inv3}), then $E$ is given by
Eq.~(\ref{pos3}) for some probability measure $\rho$ in $\R^3$. If
$\varphi \in C_c\left( \R^3 \right)$, let
\begin{equation*}
E(\varphi)=\int_{\R^3} \varphi (\vec{x}) dE(\vec{x}).
\end{equation*}
For all $f\in L^2\left( \R^3 \right)$, define the measure
\begin{equation*}
d\mu _f (\vec{x}) = \left| f(\vec{x}) \right|^2 d\vec{x}.
\end{equation*}
We then have
\begin{equation*}
\ip{f}{E(\varphi)f} = \left( \mu _f * \rho \right) (\varphi).
\end{equation*}
From (\ref{rot3}) it then follows
\begin{equation} \label{covmis}
\left( \mu _{D(R)f} * \rho \right) (\varphi) = \left( \mu _{f} *
\rho \right) \left(R^{-1}\cdot \varphi \right),
\end{equation}
where $\left(R^{-1}\cdot \varphi \right) (\vec{x}) = \varphi
(R\vec{x})$ $\forall \vec{x} \in \R^3$. Rewriting explicitly
(\ref{covmis}), we then find
\begin{equation*}
\begin{split}
& \int_{\R^3\times \R^3} \varphi (\vec{x}+\vec{y}) \left| f\left(R^{-1}\vec{x}\right)
\right|^2 d\vec{x} d\rho(\vec{y}) \\
& = \int_{\R^3\times \R^3} \left(R^{-1}\cdot \varphi \right)
\left(\vec{x}+\vec{y}\right) \left| f\left(\vec{x}\right) \right|^2 d\vec{x} d\rho(\vec{y}),
\end{split}
\end{equation*}
With some computations, setting $\psi\left(\vec{x}\right) = \left(R^{-1}\cdot \varphi \right)
\left(-\vec{x}\right)$,
this gives
\begin{equation} \label{questa}
\int_{\R^3} \left(\psi * \left| f \right|^2\right)(-\vec{y})
d\left(R^{-1}\cdot \rho\right)(\vec{y}) =
\int_{\R^3} \left(\psi * \left| f \right|^2\right)(-\vec{y})
d\rho (\vec{y}).
\end{equation}
Letting $\psi$ and $f$ vary in $C_c\left( \R^3 \right)$,
the functions $\psi * \left| f \right|^2$ span a dense subset
of $C_0\left( \R^3 \right)$. From Eq.~(\ref{questa}), it then
follows that $R^{-1}\cdot \rho = \rho$.
\end{proof}

\begin{proposition}
Let $E$ be a position observable on $\R^3$. The following facts are
equivalent:
\begin{itemize}
\item[(a)] $\no{E(U)}=1$ for every nonempty open set $U\subset\R$;
\item[(b)] $E$ is a sharp observablen ;
\item[(c)] $E=\Pi_{\vec{Q}}$.
\end{itemize}
\end{proposition}

\begin{proof}
It is clear that (c)$\Rightarrow$(b)$\Rightarrow$(a). Hence, it is
enough to show that (a) implies (c). As in the proof of Proposition
\ref{dilat}, it follows from (a) that $\rho=\delta_{\vec{t}}$ for some
$\vec{t}\in\R^3$. However, the probability measure $\delta_{\vec{t}}$ is
rotation invariant if and only if $\vec{t}=\vec{0}$. This means that $E=\Pi_{\vec{Q}}$.
\end{proof}

\section{Appendix}

\subsection{Translation covariant and boost invariant observables
in dimension $n$}\label{proof1}
Let $N=\mathbb{R}^{n+1}$ and $H=\mathbb{R}^n$, with the usual
structure of additive Abelian groups. Denote with $(p,t)$, $p\in \R^{n}$, $t\in \R$, an element of
$N$. Let $H$ act on $N$ as
\begin{equation*}
\alpha _{q}\left( p,t\right) =\left( p,t+q\cdot p\right) \quad q\in H,\
\left( p,t\right) \in N\text{.}
\end{equation*}
The Heisenberg group is the semidirect product $G=N\times_{\alpha }H$
 (see ~\cite{Heisenberg}). We will denote an element $nq \in G$, with $n = (p,t)\in N$ and
$q\in H$, as $\left((p,t),q\right)$.

Let $W$ be the following irreducible unitary representation of $G$
acting in
$L^{2}\left( \mathbb{R}^{n}\right) $
\begin{equation*}
\left[ W\left( (p,t),q\right) f\right] \left( x\right) =e^{-i\left( t-p\cdot x\right) }
f\left( x-q\right) \text{.}
\end{equation*}
Clearly, $W\left( (0,0),q\right) =U\left( q\right) $, $W\left(
(p,0),0\right) =V\left( p\right) $, and
$W\left( (0,t),0\right) =e^{-it}$. The groups $H$ and $G/N$ are
naturally identified. With such an identification, the canonical
projection $\pi :G\longrightarrow G/N$ is
\begin{equation*}
\pi \left( (p,t),q\right) =q\text{,}
\end{equation*}
and an element $\left( (p,t),q\right) \in G$ acts on $q_{0}\in H$ as
\begin{equation*}
\left( (p,t),q\right) \left[ q_{0}\right] =\pi \left( \left( (p,t),q\right)
\left( (0,0),q_{0}\right) \right) =q+q_{0}\text{.}
\end{equation*}
An observable $E$ based on $\mathbb{R}^n$ and acting in $L^{2}\left( \mathbb{R}^n
\right) $ then satisfies the analogues of Eqs.~(\ref{cov}),~(\ref{inv}) in dimension
$n$ if, and only if, for all $X\in \mathcal{B}\left( \mathbb{R}\right) $ and
$\left( (p,t),q\right) \in G$,
\begin{equation}
W\left( (p,t),q\right) E\left( X\right) W\left( (p,t),q\right)^{\ast}=E\left( X+q\right)
 \text{,}  \label{problema}
\end{equation}
i.e.~if and only if $E$ is a $W$-covariant observable based on
$G/N$. By virtue
of the Generalized Imprimitivity Theorem (see Refs.~\cite{CD},~\cite{Cattaneo}), $E$ is
$W$-covariant if and only if there exists a representation $\sigma $
of $N$ and an isometry $L$ intertwining $W$ with the induced
representation ${\rm ind}_{N}^{G}\left( \sigma \right) $ such that
\begin{equation*}
E\left( X\right) =L^{\ast }P\left( X\right) L
\end{equation*}
for all $X\in \mathcal{B}\left( \mathbb{R}^n \right) $, where $P$ is
the canonical projection valued measure of the induced
representation. Since ${\rm ind}_{N}^{G}\left( \sigma \right)
\subset {\rm ind}_{N}^{G}\left( \sigma' \right)$ (as
imprimitivity systems) if $\sigma
\subset \sigma'$ (as representations),
it is not restrictive to assume that such
$\sigma $ has constant infinite multiplicity, so that there exists
a positive Borel measure $\mu _{\sigma }$ on $\widehat{N}=
\mathbb{R}^{n+1}$ and an infinite dimensional Hilbert space
$\mathcal{H}$ such that $\sigma $ is the diagonal representation
acting in $L^{2}\left( \mathbb{R}^{n+1},\mu _{\sigma};\mathcal{H}\right)$, i.~e.
\begin{equation*}
\left[ \sigma \left( p,t\right) \phi \right] \left( h,k\right)
=e^{ih\cdot p}e^{ikt}\phi \left( h,k\right) \text{.}
\end{equation*}
Denote with $\gamma_{h,k}$, $h\in \R^n$, $k\in \R$ the following character of $N$
\begin{equation*}
\gamma_{h,k}(p,t)= e^{ih\cdot p}e^{ikt}.
\end{equation*}
The action of $H$ on $\widehat{N}$ is given by
\begin{equation*}
\left(q\cdot\gamma_{h,k}\right)(p,t)=\gamma_{h,k}\left(\alpha_{-q}
\left(p,t\right)\right)= e^{i(h-kq)\cdot p}e^{ikt},
\end{equation*}
or in other words
\begin{equation*}
q\cdot\gamma_{h,k}=\gamma_{h-kq,k}.
\end{equation*}
If $k\neq 0$, the orbit passing through $\gamma_{h,k}$ is
\begin{equation*}
\mathcal{O}_{\gamma_{h,k}}=\mathbb{R}^n \times \{k\}
\end{equation*}
and the corresponding stability subgroup is
\begin{equation*}
H_{\gamma_{h,k}}= \{0\}.
\end{equation*}
From the Mackey Machine it follows that the representations
\begin{equation*}
\rho_{h,k}:= {\rm ind}_N^G\left(\gamma_{h,k}\right)
\end{equation*}
are irreducible if $k\neq 0$, $\rho_{h,k}$ and $\rho_{h',k'}$ are
inequivalent if $k\neq k'$ and, fixed $k\neq 0$, $\rho_{h,k}$ and
$\rho_{h',k}$ are equivalent.

The representation $\rho:= {\rm ind}_N^G(\sigma)$ acts on
$L^2\left(\mathbb{R}^n, dx ; L^{2}\left( \mathbb{R}^{n+1},
\mu_{\sigma};\mathcal{H}\right)\right)$ according to
\begin{equation*}
\left[\rho\left( (p,t),q\right)f\right]\left(x\right)=
\sigma\left(p,t-p \cdot x\right)f\left(x-q\right).
\end{equation*}
Using the fact that $\sigma$ acts diagonally in
$L^{2}\left( \mathbb{R}^{n+1},\mu _{\sigma};\mathcal{H}\right)$ and the
identification  $L^2\left(\mathbb{R}^n,dx;L^{2}\left( \mathbb{R}^{n+1},
\mu_{\sigma};\mathcal{H}\right)\right)\cong
L^2\left(\mathbb{R}^n\times \mathbb{R}^{n+1},dx\otimes d\mu_\sigma (x);
\mathcal{H}\right)$, we
find that $\rho$ acts on $L^2\left(\mathbb{R}^n\times \mathbb{R}^{n+1},dx\otimes d
\mu_\sigma (x);\mathcal{H}\right)$ as
\begin{equation*}
\left[\rho\left((p,t),q\right)f\right]\left(x,h,k\right)=
e^{ih\cdot p}e^{ik\left(t-p\cdot x\right)}f\left(x-q, h,
k\right).
\end{equation*}
Write $\mu_\sigma=\mu_{\sigma_1}+\mu_{\sigma_2}$, where
$\mu_{\sigma_1}\perp \mu_{\sigma_2}$ and $\mu_{\sigma_2}
\left(\mathcal{O}_{\gamma_{0,-1}}\right)=0$, and let $\sigma=
\sigma_1\oplus\sigma_2$ be the corresponding decomposition of
$\sigma$. We then have
\begin{equation*}
{\rm ind}_N^G(\sigma)={\rm ind}_N^G(\sigma_1)\oplus {\rm ind}_N^G(\sigma_2),
\end{equation*}
where the two representations in the sum are disjoint and the sum
is a direct sum of imprimitivity systems. Since $W\simeq {\rm ind}_N^G
\left( \gamma_{0,-1} \right)$, it is not restrictive to
assume $\sigma=\sigma_1$, i.e.~that $\mu_{\sigma}$ is
concentrated in the orbit
$\mathcal{O}_{\gamma_{0,-1}}=\mathbb{R}^n\times
\{-1\}\cong \mathbb{R}^n$. Let $T$ be the following unitary operator
in $L^2(\mathbb{R}^n\times \mathbb{R}^n,dx\otimes d\mu_\sigma (x)
;\mathcal{H})$:
\begin{equation*}
[Tf]\left(x ,h\right)=f\left(x + h  ,h\right).
\end{equation*}
If we define the representation $\hat{\rho}$, given by
\begin{equation*}
\left[\hat{\rho}\left((p,t),q\right)f\right]\left(x,h\right)=
e^{-i\left(t-p\cdot x\right)}f\left(x-q, h\right),
\end{equation*}
then $T$ intertwines $\hat{\rho}$ with $\rho$. Since
$\hat{\rho}\simeq W\otimes I_{L^2\left(\mathbb{R}^n,\mu_\sigma;
\mathcal{H}\right)}$ and $W$ is irreducible, every isometry
intertwining $W$ with $\hat{\rho}$ has the form
\begin{equation*}
[\widetilde{L}f]\left(x ,h\right)=f\left(x \right) \varphi (h) \quad
\forall f \in L^2(\mathbb{R}^n)
\end{equation*}
for some $\varphi \in L^2(\mathbb{R}^n,\mu_\sigma ;\mathcal{H})$
with $\left\|\varphi \right\|_{L^2} = 1$. The most general
isometry $L$ intertwining $W$ with $\rho$ has then the
form $L = T\widetilde{L}$ for some choice of $\varphi$, and the
corresponding observable is given by
\begin{eqnarray*}
\ip{g}{E(X)f} &=& \ip{g}{L^*P(X) Lf} =
\ip{T\widetilde{L}g}{P(X)T\widetilde{L}f} \\
&=& \int_{\mathbb{R}^{2n}}\chi_X (x)f(x+h) \overline{g(x+h)}
\left\langle \varphi (h),\varphi (h) \right\rangle dx d\mu_\sigma
(h).
\end{eqnarray*}
It follows that
\begin{eqnarray*}
\left[ E(X)f \right] (x) &=& f(x) \int_{\mathbb{R}^n}\chi_X (x-h)
\left\|
\varphi (h) \right\|^2 d\mu_\sigma (h) \\
&=& f(x) \int_{\mathbb{R}^n}\chi_X (x-h) d\mu (h),
\end{eqnarray*}
where $d\mu (h) = \| \varphi (h)\|^2 d\mu_\sigma (h)$
is a probability measure on $\mathbb{R}^n$.

\subsection{Supplement to Remark \ref{r_dilation}}\label{supplement}

Let $A'(a) = U(t)A(a)U(t)^*$. Then, $A'(a)\Pi_Q (X) A'(a)^* = \Pi_Q (aX)$. Denote with
$\Pi^+_Q$ the restriction of $\Pi_Q$ to the Borel subsets of $\R_+$.
Then, $S_0 = \left(A_0, \Pi^+_Q,
L^2\left(0,+\infty\right)\right)$ and $S = \left(A', \Pi^+_Q,
L^2\left(0,+\infty\right)\right)$ are transitive imprimitivity systems
of the group $\R_+$ based on $\R_+$.
Using the Mackey Imprimitivity Theorem, there exists a Hilbert space $\mathcal{K}$
such that $S = {\rm ind}_{\{1\}}^{\R_+} (I_{\mathcal{K}})$, where
$I_{\mathcal{K}}$ is the trivial representation of ${\{1\}}$
acting in $\mathcal{K}$. Since $S_0 = {\rm ind}_{\{1\}}^{\R_+} (1)$,
we have the isomorphism of intertwining operators
$\mathcal{C} \left(1,I_\mathcal{K}\right) \simeq
\mathcal{C} \left(S_0,S\right)$, and hence there exists an isometry
$W_1 : L^2\left(0,+\infty\right) \longrightarrow
L^2\left(0,+\infty\right)$ intertwining $S_0$ with $S$.
In particular, $W_1 \Pi^+_Q = \Pi^+_Q W_1$, and hence
there exists a measurable function $\beta_1 : \R_+ \longrightarrow
\mathbb{T}$ such that
\begin{equation*}
[W_1 f](x)=\beta_1 (x) f(x)\quad \forall f\in L^2\left(0,+\infty\right).
\end{equation*}
It follows that $W_1$ is unitary. Reasoning as above, one finds a unitary operator $W_2$
intertwining the restrictions of $A_0$ and $A'$ to $L^2\left(-\infty , 0\right)$, with
\begin{equation*}
[W_2 f](x)=\beta_2 (x) f(x)\quad \forall f\in L^2\left(-\infty , 0\right),
\end{equation*}
for some measurable function $\beta_2 : \R_- \longrightarrow
\mathbb{T}$.
Then, $\widetilde{W}=W_1\oplus W_2$ is unitary on
$L^2\left(-\infty,+\infty\right)$, and $A(a) = U(t)^* \widetilde{W}A_0 (a) \widetilde{W}^* U(t)$
has the claimed form for all $a\in \R_+$.

\section*{Acknowledgments}

The authors wish to thank Gianni Cassinelli, Pekka Lahti and Kari
Ylinen for discussions on this topic and their comments on this paper.

\end{document}